\documentclass[prd, showpacs, twocolumn, floatfix]{revtex4}

\usepackage{graphicx}
\usepackage{amsmath, amsfonts, amssymb, bm}
\usepackage{hyperref}

\usepackage{slashed}
\usepackage[mathscr]{euscript}

\newcommand{\cf}{\mathscr{X}_{[0,2\pi]}}
\newcommand{\pv}{\phi}
\newcommand{\Nosc}{N_{\text{osc}}}
\newcommand{\cep}{\chi}
\newcommand{\xim}{\xi_\text{max}}

\newcommand{\RKG}[1]{\mathscr{R}^{\text{KG}}_{#1}}
\newcommand{\RFlip}[1]{\mathscr{R}^{(1)}_{#1}}
\newcommand{\RNoFlip}[1]{\mathscr{R}^{(0)}_{#1}}
\newcommand{\Obeta}[1]{O(\beta^{#1})}

\newcommand{\DiKg}{\zeta}

\begin{document}

\title{Strong-field Breit-Wheeler pair production in short laser pulses:\\
Relevance of spin effects}

\author{M. J. A. Jansen$^1$}
\author{J. Z. Kami\'nski$^2$}
\author{K. Krajewska$^2$}
\author{C. M\"uller$^1$}
\affiliation{$^1$Institut f\"ur Theoretische Physik I, Heinrich-Heine-Universit\"at D\"usseldorf, Universit\"atsstr. 1, 40225 D\"usseldorf, Germany\\
$^2$Institute of Theoretical Physics, Faculty of Physics, University of Warsaw, Pasteura 5, 02-093 Warsaw, Poland}

\date{\today}

\begin{abstract}
Production of electron-positron pairs in the collision of a high-energy photon with a high-intensity few-cycle laser pulse is studied. By utilizing the frameworks of laser-dressed spinor and scalar quantum electrodynamics, a comparison between the production of pairs of Dirac and Klein-Gordon particles is drawn. Positron energy spectra and angular distributions are presented for various laser parameters. We identify conditions under which predictions from Klein-Gordon theory either closely resemble or largely differ from those of the proper Dirac theory. In particular, we address the question to which extent the relevance of spin effects is influenced by the short duration of the laser pulse. 
\end{abstract}
 
\maketitle

\section{Introduction}

Very strong electromagnetic fields can extract electron-positron ($e^-e^+$) pairs from the quantum vacuum. With respect to the strong fields provided by a high-intensity laser wave, the first theoretical studies on $e^-e^+$ pair production via multiphoton absorption date back to the 1960s \cite{Reiss,Ritus}. Due to a remarkable and still ongoing progress in high-power laser technology, which has started in the mid 1980s, there are clear prospects for corresponding experimental studies in the near future. Optical field intensities well above 10$^{20}$~W/cm$^2$ are routinely available today in many laboratories worldwide and an increase towards 10$^{25}$~W/cm$^2$ is envisaged \cite{ELI, XCELS}. This way the characteristic intensity level for vacuum pair production -- given by the critical value $I_{\rm cr}\sim 10^{29}$~W/cm$^2$ -- is being approached. Besides, high-intensity x-ray beams can be generated nowadays at free-electron laser facilities \cite{XFEL} and through plasma harmonics \cite{surfaceHHG}.

With the aid of a suitable combination of advanced technologies, experimental studies on $e^-e^+$ pair production in strong laser fields are feasible already at present. The relevant field frequency and intensity can be enhanced effectively when a high-energy photon or particle beam counterpropagates an intense laser pulse. This kind of setup has enabled the first observation of $e^-e^+$ pair creation via multiphoton absorption.  It was accomplished in ultrarelativistic electron-laser collisions at the Stanford Linear Accelerator Center (SLAC) in the 1990s \cite{SLAC}. The detected pairs were attributed to the reaction 
\begin{eqnarray}
\label{BW}
\omega_\gamma + N\omega_L \to e^-e^+\ ,
\end{eqnarray}
involving a high-energy photon $\omega_\gamma$ generated through Compton backscattering and a certain number $N$ of laser photons $\omega_L$. Equation \eqref{BW} represents a multiphoton version of the well-known Breit-Wheeler process \cite{BW}.

The successful SLAC experiment together with the promising prospects for a new generation of ultra-high intensity laser laboratories have stimulated considerable theoretical activities on $e^-e^+$ pair production and other processes of quantum electrodynamics (QED) in very strong laser fields during the last decade \cite{review1, review2}. A special focus has been placed on strong-field QED calculations in laser fields of finite extent because very high intensities are reached in short laser pulses, comprising just a few field oscillations. Corresponding studies have been carried out on multiphoton Compton scattering \cite{Boca,Seipt,Mackenroth,RoshchupkinCompton,KasiaCompton,Harvey,Akal} where characteristic imprints of the finite pulse shape on the scattered photon distribution were shown. Also the multi\-photon Breit-Wheeler process \eqref{BW} in a finite laser pulse has been examined \cite{Fedorov,Heinzl,Kaempfer,KasiaBW,RoshchupkinBW,Fedotov,Huayu, Meuren,Jansen,zepto}, which is related to multi\-photon Compton scattering by a crossing symmetry of the corresponding Furry-Feynman graphs. Here, the frequency spectrum of the pulse is reflected in the resulting electron and positron momenta. Similar effects were obtained for $e^-e^+$ pair production by the strong-field Bethe-Heitler process involving multiphoton absorption in the presence of a nuclear Coulomb field \cite{RoshchupkinBH,KasiaBH}. The influence of the pulse shape was also studied for pair production processes in strong electric fields of finite temporal or spatial extension \cite{Dunne,Grobe}.

Another aspect of particular interest in strong-field processes is the relevance of spin effects. They have been studied in laser-atom interactions, such as relativistic photoionization \cite{Faisal,Klaiber} and high-harmonic generation \cite{HHG}, as well as in high-intensity QED phenomena. One possibility to analyze spin effects in relativistic processes in strong laser fields is to compare the predictions from the proper spinor QED based on the Dirac equation with those from the scalar theory based on the Klein-Gordon equation where the electron spin is neglected. Corresponding comparative studies have been carried out with respect to Compton scattering \cite{ComptonSpin}, Mott scattering \cite{Mott}, Kapitza-Dirac scattering \cite{Sven}, and Bethe-Heitler pair production \cite{Tim-Oliver}. For strong-field Breit-Wheeler pair production in an intense monochromatic laser wave, spin-resolved results were obtained within Dirac theory \cite{Tsai,Ivanov} and a comparison of total production rates for spin-$\frac{1}{2}$ and scalar particles was performed based on the laser-dressed polarization operator \cite{Selym}. Characteristic differences between fermions and bosons are also known from pair production in oscillating electric fields, where the location of Rabi-like resonances depends on the particle quantum statistics \cite{Popov}. The latter is crucial for the Klein paradox which has recently been examined within a computational field-theoretical approach; pronounced suppression and enhancement effects were revealed for Dirac and Klein-Gordon particles, respectively \cite{Klein}. In standing laser waves of elliptical polarization,  spin-polarized $e^-e^+$ pairs may be produced \cite{Bauke}. We note that very recently, spin effects in multiphoton Compton scattering in a pulsed laser field were analyzed \cite{KasiaComptonSpin}.

In the present paper, we study strong-field Breit-Wheeler pair production in finite laser pulses, focussing on spin effects. Production probabilities for pairs of Dirac and Klein-Gordon particles are obtained within the frameworks of laser-dressed spinor and scalar QED, respectively. By way of comparison, the relevance of spin effects in the process is revealed by inspecting positron energy spectra and angular distributions. Also spin-resolved calculations are carried out. Our main goal is to reveal under which conditions spin effects are either more or less pronounced in a short laser pulse as compared with an infinitely extended laser wave.

Our paper is organized as follows. In Sec.~II we briefly survey the theoretical frameworks of strong-field Breit-Wheeler pair production in short laser pulses within spinor and scalar QED, respectively. A benchmark for the subsequent discussion of spin effects in pulsed fields is established in Sec.~III by providing analytical expressions for total Breit-Wheeler pair production rates in monochromatic laser fields. Their main properties are discussed and their understanding is supported by an intuitive picture. In Sec.~IV we present our numerical results for positron energy spectra and angular distributions in few-cycle laser pulses. Various parameter combinations are considered allowing us to identify conditions under which the relevance of spin effects is enhanced or reduced due to the short duration of the laser pulse. 
As we shall demonstrate, the impact of the finite pulse length and shape can be surprisingly large.
We finish with concluding remarks in Sec.~V.

\section{Theoretical Framework}
In this section, we first introduce the required definitions and notation and then present the analytical derivation of the particle creation probability for scalar particles obtained in the collision of a high-energy gamma quantum and a short laser pulse.
The analogous derivation for Dirac particles was carried out in \cite{Jansen} where a similar notation is used. More details can also be found in \cite{KasiaBW}.

\subsection{Definitions}
Relativistic units with $\hbar=c=1$ shall be used, unless explicitly stated otherwise.
The positron charge and mass are denoted by $e$ and $m$, respectively.
We employ the metric tensor $\operatorname{diag}(+,-,-,-)$, so that the four-product of two four-vectors $a^\mu = (a^0,{\bf a})$ and $b^\mu = (b^0,{\bf b})$ reads $a\cdot b = a^0b^0-{\bf a}\cdot{\bf b}$. Feynman slash notation is used to denote four-products with Dirac $\gamma$-matrices.

\subsubsection{Laser pulse}
The laser pulse is defined by its vector potential in radiation gauge
\begin{equation}
\label{LasVecPot}
 A^\mu = A_0 f(\pv) \cf(\pv) \epsilon^\mu
\end{equation}
with the amplitude parameter $A_0$ and the real polarization vector $\epsilon^\mu$. 

Assuming a uniform propagation direction ${\bf n}\bot{\boldsymbol\epsilon}$ for all spectral components of the pulse, the spacetime dependence is determined by the phase $\pv = k \cdot x$. Here, the fundamental wave four vector $k^\mu = \omega_b(1,{\bf n})$ with basic frequency $\omega_b$ has been introduced, as well as the spacetime coordinate $x^\mu=(t,{\bf x})$.
The pulse shape is determined by the combination of the shape function $f(\pv)$ and of the characteristic function $\cf(\pv)$. In order to model a finite pulse, the latter restricts the phase variable to the interval $[0,2\pi]$. 

\begin{figure}
 \includegraphics[width=\columnwidth]{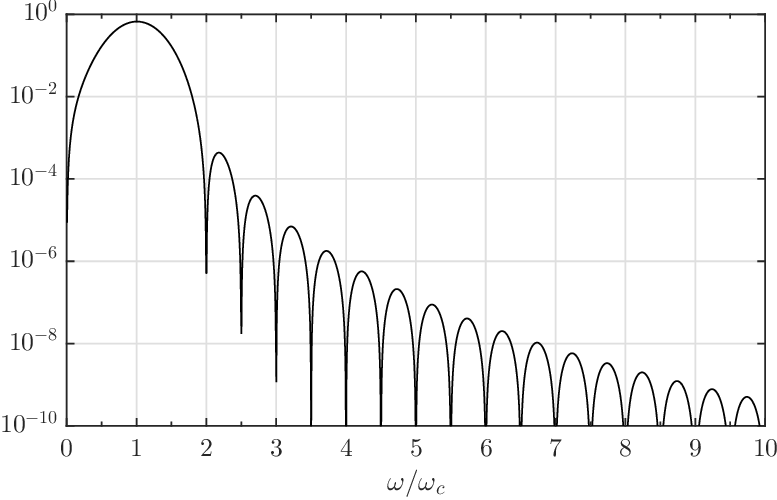}
\caption{Spectral energy density (arb. units) of the ultrashort pulse with a shape function given by Eq.~\eqref{shapefunction} with $\Nosc=2$ and $\cep=0$. The characteristic zeros at integer multiples of the basic frequency $\omega_b = \omega_c / \Nosc$ (as indicated by the subticks) are caused by the window function $\cf(\pv)$.}
\label{pulse}
\end{figure}

While we present the following derivation without specifying the actual shape, our numerical results are obtained for a specific shape which is defined by means of its derivative
\begin{equation}
\label{shapefunction}
 f^\prime(\pv) = \sin^2 \left(\pv/2 \right) \sin(\Nosc \pv+\cep)
\end{equation}
where  $\Nosc$ gives the number of oscillations within the $\sin^2$ - envelope, and $\cep$ allows to vary the carrier-envelope phase.
The spectrum, as presented in Fig.~\ref{pulse}, is dominated by a broad peak at the central frequency $\omega_c = \Nosc \omega_b$. Especially for higher energies, the window function $\cf(\pv)$ causes a series of spectral holes at integer multiples of the laser basic frequency. As we shall see later on, this characteristic structure is reflected in the energy spectra of the produced particles.

We measure the field strength by means of the Lorentz-invariant parameter
\begin{equation}
 \xim = \frac{eA_0}{m} \text{max}_\pv |f(\pv)|\,.
\end{equation}

\subsubsection{Volkov States}
\label{Sec:Volkov_States}
Both the Klein-Gordon equation and the Dirac equation can be solved in closed analytical form for a particle being subject to a plane-wave fronted pulse as given by Eq. \eqref{LasVecPot} by means of the Volkov states \cite{IZ,LL}. 

For a scalar particle (anti-particle) with free four-momentum $p^\mu_-$ ($p^\mu_+$), the Gordon-Volkov states read
\begin{equation}
 \Phi_{p_\pm} = \frac{1}{\sqrt{2VE_{p_\pm}}} e^{i[\pm p_\pm \cdot x + \Lambda_\pm]}
\end{equation}
with 
\begin{equation}
\Lambda_\pm = \frac{1}{k \cdot p_\pm}\int^{k\cdot x}_0 \left[ e p_\pm \cdot A(\pv) \mp \frac{e^2}{2}  A^2(\pv) \right] d\pv
\end{equation}
and $p^\mu_\pm = (E_{p_\pm},{\bf p}_\pm)$, $E_{p_\pm} = \sqrt{m^2 + {\bf p}_\pm^2}$  and a normalizing volume $V$.

Conversely, for a spin-$\frac{1}{2}$ particle, the Dirac-Volkov states read
\begin{equation}
 \Phi_{p_\pm}^{(1/2)} = \sqrt{\frac{m}{VE_{p_\pm}}} \left[ 1 \pm \frac{e \slashed{k}\slashed{A}}{2 k \cdot p_\pm} \right] w_{\pm} e^{i\left[\pm p_\pm \cdot x+\Lambda_\pm\right]} \,.
\end{equation}
The free spinors $w_{\pm}$ satisfy the algebraic equation $(\slashed{p} \pm m) w_{\pm} = 0$ and are normalized according to $\overline{w}_{\pm}(p_\pm,s_\pm) w_{\pm}(p_\pm,s^\prime_\pm) = \mp \delta_{s_\pm,s^\prime_\pm}$ and $\overline{w}_{\pm}w_{\mp}=0$,
where $s_\pm$ labels the spin projection of the positron and electron, respectively \cite{Greiner}.
The spin quantization axis is chosen along the propagation direction of the laser pulse.

\subsubsection{Gamma Quantum}
The pair production process in the strong-field Breit-Wheeler scenario is induced by the decay of a high-energy gamma quantum traveling in the strong laser field.
This gamma quantum is described as a single mode of a quantized radiation field.
Its absorption during the process gives rise to an effective scattering potential
\begin{eqnarray}
  A_\gamma^\mu=\sqrt{\frac{2 \pi}{V \omega_\gamma}}\, e^{-i k_\gamma \cdot x}\, \epsilon_\gamma^\mu \,,
\end{eqnarray}
where $k_\gamma^\mu = (\omega_\gamma,{\bf k}_\gamma)$ is the corresponding wave four vector whereas $\epsilon_\gamma^\mu$ is a real polarization four vector. To simplify our calculations, the gamma quantum is assumed to be colliding head-on with the laser pulse.

\subsection{Pair Creation Probability}
In the following, we present the derivation of the pair creation probability for Klein-Gordon particles in a short laser pulse. 
Within laser-dressed scalar QED, the $S$-matrix element for the creation of a (spinless) electron-positron pair with momenta $p_-^\mu$ and $p_+^\mu$ reads
\begin{eqnarray}
 \mathcal{S}_{p_+p_-}= -i \int  d^4x\,  \Phi_{p_-}^*  \operatorname{H}_{\rm{int}}  \Phi_{p_+} ,
\end{eqnarray}
with the interaction Hamiltonian
\begin{eqnarray}
\operatorname{H}_{\rm{int}}=-ie \left( {\bf A}_\gamma \cdot \overset{\rightarrow}{\boldsymbol{\nabla}} - \overset{\leftarrow}{\boldsymbol{\nabla}} \cdot {\bf A}_\gamma \right) + 2e^2 {\bf A} \cdot {\bf A}_\gamma \,.
\end{eqnarray}
The $S$ matrix can be brought into the form
\begin{equation}\label{Smatrix}
 \mathcal{S}_{p_+p_-} =S_0 \int d^4x \, C(\pv) e^{-iQ\cdot x-iH(\pv)}
\end{equation}
with $S_0 = -iem \sqrt{ \frac{\pi}{ 2V^3 E_{p_+} E_{p_-} \omega_\gamma }}$ and
\begin{equation}
 \begin{split}
 C(\pv) &= g_0 + g_1 f(\pv)\cf(\pv) \,,\\
H(\pv) &= \int_0^\pv h(\tilde{\pv}) d\tilde{\pv} \,,\\
 Q^\mu &= k_\gamma^\mu - \left( p_+^\mu + p_-^\mu\right) \,,
 \end{split}
\end{equation}
and with abbreviations
\begin{equation}
 \begin{split}
g_0 &=  \frac{ {\bf p}_- - {\bf p}_+ }{m}  \cdot {\boldsymbol\epsilon}_\gamma \,,\\
g_1 &=  \frac{2eA_0}{m}  {\boldsymbol\epsilon} \cdot {\boldsymbol\epsilon}_\gamma \,,\\
h({\pv}) &= \left[ h_1 f({\pv}) + h_2 f^2({\pv})\right] \cf(\pv) \,,\\
  h_1 &= -eA_0 \left[ \frac{\epsilon \cdot p_+}{k \cdot p_+} - \frac{\epsilon \cdot p_-}{k \cdot p_-} \right] \,,\\
h_2 &=- \frac{e^2A_0^2}{2} \left[ \frac{1}{k\cdot p_+}+\frac{1}{k \cdot p_-}\right] \,.
 \end{split}
\end{equation}

In order to carry out the spacetime integration, the term proportional to $g_0$ is transformed according to (see also \cite{KasiaCompton})
\begin{equation}
 g_0 \rightarrow \frac{-k^0}{Q^0} h(\pv) g_0 \,.
\end{equation}
The condition $Q^0\neq 0$ follows from kinematical constraints \cite{KasiaBW}. 
Introducing light-cone coordinates, three integrations can be carried out directly.
For a given four vector $x^\mu$ and with $x^\parallel = {\bf x}\cdot{\bf n}$, we introduce $x^- = x^0 - x^\parallel$, $x^+ = \frac{1}{2}\left(x^0+x^\parallel\right)$, and ${\bf x}^\bot = {\bf x} - x^\parallel {\bf n}$. 
We obtain
\begin{equation}\label{SMatrixLastIntegration}
\begin{split}
 \mathcal{S}_{p_+p_-} = &(2\pi)^3 S_0 \delta(Q^-)\delta^{(2)}({\bf Q}^\bot) \\
&\times \int_0^{2\pi/k^0} dx^- C(k^0x^-)e^{-iQ^0x^- -iH(k^0x^-)} \,.
\end{split}
\end{equation}
The remaining integral can be calculated numerically.
The total creation probability of Klein-Gordon (KG) pairs is obtained from
\begin{equation}
 \mathscr{P}^{\text{KG}} = \frac{1}{2}\sum_{\lambda_\gamma} \int \frac{Vd^3p_+}{(2\pi)^3} \int \frac{Vd^3p_-}{(2\pi)^3} |\mathcal{S}_{p_+p_-} |^2 \,,
\end{equation}
assuming an unpolarized beam of gamma quanta.

Comparing the derivation above with the spin-$\frac{1}{2}$ case, where the $S$ matrix reads
\begin{eqnarray}
 \mathcal{S}_{p_+s_+,p_-s_-}^{(1/2)}= ie \int  d^4x\,  \overline{\Phi}_{p_-s_-}^{(1/2)}  \slashed{A}_\gamma \Phi_{p_+s_+}^{(1/2)} ,
\label{S-Dirac}
\end{eqnarray}
we note that the expressions following Eq. \eqref{Smatrix} are mostly equivalent, except for the prefactor $S_0$ and the matrix elements $g_0$ and $g_1$ (see Eqs.~(10) and (12) in \cite{Jansen}). The latter include the spinor properties of the Dirac particles. The total production probability in the Dirac case can be decomposed into spin contributions,
\begin{eqnarray}
\label{P-Dirac}
\mathscr{P}^{(1/2)} = \sum_{s_+,s_-} \mathscr{P}_{s_+s_-}^{(1/2)}\ ,
\end{eqnarray}
where the quantitity $\mathscr{P}_{s_+s_-}^{(1/2)}$ is given by Eq.~(16) with $\mathcal{S}_{p_+p_-}$ being replaced by $\mathcal{S}_{p_+s_+,p_-s_-}^{(1/2)}$. We note that, in general, the formal quantitity $\mathscr{P}_{s_+s_-}^{(1/2)}$ cannot be interpreted straightforwardly as the probability to create a positron and an electron with spin projections $s_+$ and $s_-$ respectively. Nevertheless, the spin decomposition in Eq.~\eqref{P-Dirac} will prove useful for gaining an intuitive understanding of the relation between the production of spinor and scalar pairs. For distinguishing the two inequivalent spin configurations in our case we introduce besides the label $s=|s_+ + s_-|$. It gives the absolute magnitude of the sum of the spin quantum numbers occuring in $\mathscr{P}_{s_+s_-}^{(1/2)}$ and has, accordingly, the possible values $s=0$ and $s=1$.

\section{Spin-resolved multiphoton processes in monochromatic fields}
The pair creation in a laser pulse of moderate intensity can be understood in terms of multiphoton processes induced by the spectral components of the pulse \cite{Kaempfer,Jansen}.
Therefore, we start our analysis of spin effects by analyzing spin-dependent pair creation rates for multiphoton processes in a monochromatic laser field.
The spin effects obtained in a short laser pulse (see Sec.~IV) will be explicable in terms of these monochromatic rates, combined with the spectral composition of the pulse.

The monochromatic laser field is assumed to be linearly polarized. Its field strength is determined by means of the usual dimensionless amplitude $\xi = \frac{eA_0}{m}$ which is assumed to be small, $\xi\ll 1$. 
We shall compare perturbative multiphoton pair production rates $\mathscr{R}_N$ in leading order of $\xi$. We present these rates, throughout, in units of a common prefactor $\alpha m \xi^{2N}$ which includes the perturbative intensity scaling for a process involving $N$ laser photons.

For notational and calculational simplicity, the rates are considered in the center-of-mass frame. The energy of both the particles and the photons can thus be measured by means of the reduced velocity parameter $\beta$ of the particles (see \cite{Jansen_Proc} for further details).
We first consider processes close to the energy threshold ($\beta=0$). This consideration will help us to develop a basic physical understanding of the influence of the particles' spin.

For the one-photon process, we obtain
\begin{align}
 \RKG{1} &=  \frac{1}{8} \beta - \frac{11}{48} \beta^3 + \Obeta{5} \label{R_Taylor_1a} \,,\\
 \RNoFlip{1} &=  \frac{1}{4} \beta - \frac{1}{8} \beta^3 + \Obeta{5} \label{R_Taylor_1b} \,,\\
 \RFlip{1} &=  \frac{1}{2} \beta^3 - \frac{7}{20}\beta^5 + \Obeta{7} \label{R_Taylor_1c} \,.
\end{align}
The spin contributions to the Dirac case are distinguished by the superscript $(s)$, which has been introduced at the end of Sec.~\ref{Sec:Volkov_States}. This is, the multiphoton Dirac rate is decomposed according to  $\mathscr{R}_N^{(1/2)}=\RNoFlip{N}+\RFlip{N}$.

A comparison of Eqs.~\eqref{R_Taylor_1a}-\eqref{R_Taylor_1c} suggests a particularly simple picture for the relation between the production of Dirac and KG pairs close to the threshold of the one-photon process. In the limit of small $\beta$, the Dirac rate is entirely determined by the contribution from $s=0$ and twice as large as the rate for (intrinsically spinless) KG pairs. Thus, intuitively one may say that Dirac pairs are produced with vanishing total spin. The rate ratio of 2 coincides with the number of possible configurations being available for the ``spinless'' Dirac pairs \cite{Selym} \footnote{Note that these results have been obtained for unpolarized gamma photons. For linearly polarized gamma photons, this simple picture fails.}.

However, moving away from the threshold by increasing the photon energies, the simple correspondence breaks down. The ratio $\RNoFlip{1}/\RKG{1}$ starts to exceed the value of 2. Additionally, the contribution from $s=1$ to the Dirac rate becomes sizeable.

For the two-photon process, one finds
\begin{align}
  \RKG{2} &=  \frac{13}{48}\beta^3 - \frac{133}{160}\beta^5 + \Obeta{7} \label{R_Taylor_2a} \,,\\
  \RNoFlip{2} &=  \frac{1}{3} \beta^3 - \frac{77}{160}\beta^5 + \Obeta{7} \label{R_Taylor_2b} \,,\\
  \RFlip{2} &= \frac{1}{8}\beta - \frac{31}{48}\beta^3 +\Obeta{5} \label{R_Taylor_2c} \,.
\end{align}
Close to the threshold, the contribution from $s=1$ to the Dirac rate is domimant this time. The contribution from $s=0$ and the KG rate still behave similarly. However, they are suppressed due to the $\beta^3$-scaling. Their numerical ratio is $\frac{16}{13}$, which strikingly differs from the factor 2 found for the one-photon process.

For the three-photon process, the spin effects resemble the one-photon process again:
\begin{align}
 \RKG{3} &= \frac{9}{512}\beta - \frac{249}{1024}\beta^3 + \Obeta{5} \label{R_Taylor_3a}\,,\\
 \RNoFlip{3} &= \frac{9}{256} \beta - \frac{177}{512}\beta^3+\Obeta{5} \label{R_Taylor_3b}\,,\\
 \RFlip{3} &= \frac{93}{128}\beta^3-\frac{1155}{256}\beta^5+\Obeta{7} \label{R_Taylor_3c} \,.
\end{align}
Close to the threshold, we find $\RNoFlip{3}/\RKG{3}=2$, while the contribution from $s=1$ to the Dirac rate is suppressed.

Increasing the photon numbers further, the described behavior (in particular close to the threshold) continues to alternate between even and odd photon numbers. This was checked for photon numbers up to $N=10$. In particular, for low-energy particles being produced by an even photon number, the ratio $\RNoFlip{N}/\RKG{N}$ was found to be described by the formula $\frac{4N^2}{2N^2+2N+1}$. Note that this expression approaches the value 2 for large photon numbers. 

An intuitive understanding of the qualitative difference between even and odd laser photon numbers can be gained by considering the angular momentum balance in the process (see also \cite{Ivanov}). The incoming $N$ laser photons and the gamma quantum carry one unit of angular momentum along the beam axis each. The resulting total angular momentum is transferred to the produced particles. This constraint imposes a selection rule which depends on the parity of the number of absorbed photons.
It becomes particularly sensitive to the spin configuration when we regard processes close to the threshold \cite{Mocken}.

For example in the case of a one-photon process, the total angular momentum of the laser photon and the gamma quantum is an even number (including zero). It must be compensated by the produced particles. In the limit of small $\beta$, the angular momentum of the particles is solely composed of their spins. This offers a plausible explanation for the suppression of the contribution from $s=1$ and the dominance of the contribution from $s=0$ to the Dirac rate $\mathscr{R}_1^{(1/2)}$.
Conversely, for a two-photon process, the total angular momentum is an odd number. This favors the contribution from $s=1$ to the Dirac rate $\mathscr{R}_2^{(1/2)}$ in the limit of small $\beta$ and suppresses, in turn, the contribution from $s=0$ as well as the KG rate.

The above line of argument breaks down when the particle energies grow. Then, the angular momentum of the photons is transferred not only to the particles' spins but also to their orbital angular momenta. 
As we shall see below, the angular distributions reveal clear signatures of this selection rule. 
When the incoming angular momentum is completely compensated by the particles' spin, they can be emitted into any direction.
Conversely, a non-zero orbital angular momentum along the beam axis is most easily attained by low-energy particles when they are emitted into transverse direction.

\begin{figure}
 \includegraphics[width=0.95\columnwidth]{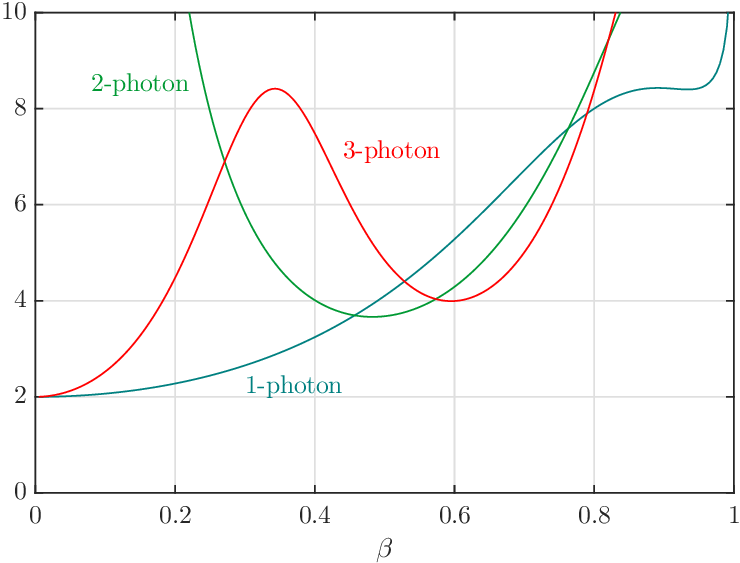}
\caption{Ratio between the rates for Dirac and KG for different numbers of absorbed photons as a function of the velocity $\beta$ of the particles.}
\label{zeta}
\end{figure}

As a preparation for the following section, we present in Fig.~\ref{zeta} the ratio $\DiKg_N=\mathscr{R}_N^{(1/2)}/\RKG{N}$ between the spin-summed Dirac and KG rates for $N=1,2,3$ in the full interval of $\beta$.
The spin effects reveal an increasingly rich dependence on $\beta$ as the number of absorbed photons grows.
While the spin sensitivity of the one-photon process essentially grows with increasing $\beta$, the ratio for the two-photon process 
diverges with $1/\beta^2$ at the threshold and falls into a minimum at $\beta\sim0.5$. The three-photon ratio $\DiKg_3$ reveals a pronounced maximum and minimum.
As a consequence, the spin sensitivities may strongly depend on the photon number at a given velocity $\beta$.
Nevertheless, we find comparable spin sensitivities $\DiKg_N\sim4$ in the range of intermediate velocities $\beta\sim0.5$.
Furthermore, since $\DiKg_N\geq2$ in the whole $\beta$ interval, the Klein-Gordon rates form a lower limit for the full Dirac rates, again with a factor of two.

In the limit of ultrarelativistic particles, the ratio $\DiKg_N$ suggests huge spin effects. Note, however, that the underlying production rates approach zero, such that these processes contribute only marginally to the full pair production probability obtained in a pulse.

\section{Spin Effects in Short Pulses}
In the following, we shall investigate spin effects in pair production driven by a short laser pulse with central frequency $\omega_c$ and draw a comparison with the corresponding process in a monochromatic laser wave of the same frequency. 
In order to have comparable amplitudes, we choose the amplitude of the monochromatic wave as $\xi = \xim$. 
The number of cycles and the carrier-envelope phase of the laser pulse are chosen as $\Nosc=2$ and $\cep = 0$ throughout.
The ratio between the probabilities (or rates) of Dirac and KG pairs is denoted by $\DiKg$. 

We shall consider three different parameter combinations which are distinguished by the value of $\omega_c \omega_\gamma$. This product determines the energy threshold of the pair production process. If $\omega_c \omega_\gamma>m_\ast^2$, with $m_\ast$ being the laser-dressed mass, a pair can be created in the collision of the gamma photon with one laser photon from a monochromatic field of frequency $\omega_c$.
The numerical calculations are carried out in a frame of reference where the gamma photon energy equals the total absorbed laser photon energy for the leading order photon-number channel.
The parameters can be achieved, for instance, by employing a Nd:YAG laser with $\omega_c\approx2.4$ eV and peak intensity $\lesssim10^{18}$ W/cm$^2$, and a high energy photon with $\sim 100$ GeV, which could be generated by Compton backscattering off an ultrarelativistic electron beam.

\subsection{Above the one-photon threshold}
We start with a laser frequency $\omega_c=1.006 m$ and a gamma quantum with equal energy. In the case of a monochromatic laser field of moderate intensity, these parameters allow a one-photon process closely above the threshold, with $\beta\approx0.11$.
For $\xi=0.02$, the monochromatic laser field yields a rate ratio of $\DiKg \approx 2.1$, in accordance with Eqs.~\eqref{R_Taylor_1a}-\eqref{R_Taylor_1c}. Employing a short two-cycle pulse, we find a substantially larger probability ratio of $\DiKg \approx 3.5$ instead.

\begin{figure}
 \includegraphics[width=\columnwidth]{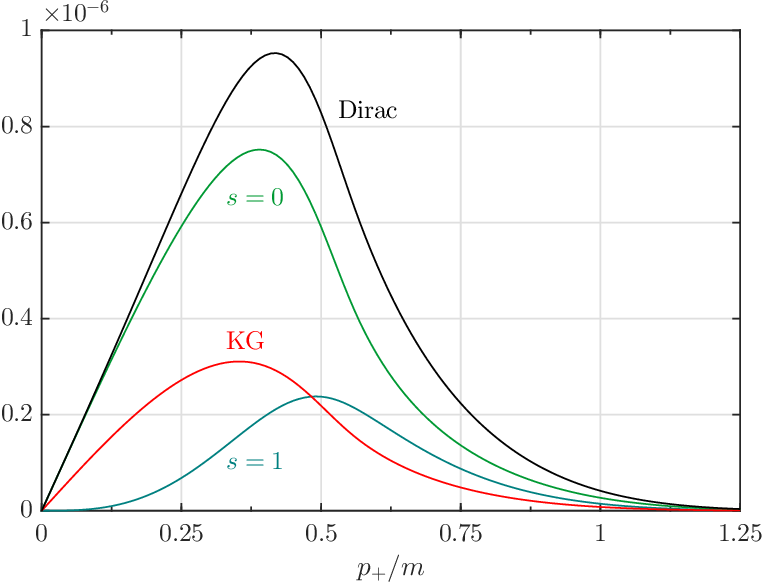}
\caption{Angularly integrated probabilities $\frac{d\mathscr{P}}{dE_{p_+}}$ in units of $1/m$ as a function of $p_+/m$ for $\omega_c=\omega_\gamma=1.006m$, $\Nosc=2$, and $\xim=0.02$.}
\label{above-energy}
\end{figure}

This enhancement of the relevance of spin effects can be attributed to the broad spectrum of the ultrashort pulse. In contrast to the monochromatic field, the pulse facilitates creation of particle pairs within a broad range of energies. As depicted in Fig.~\ref{above-energy}, the dominant pair production channels are actually those with energies well above the monochromatic case. In fact, the largest contributions stem from positron momenta around $p_+\approx 0.46m$. For the corresponding value of $\beta\approx 0.42$, the ratio $\DiKg_1\approx 3.4$ can be read off from Fig.~\ref{zeta}. These energetic positrons have been produced by laser photons with frequencies significantly above $\omega_c$. Although the spectral weight of these photons is reduced as compared with photons close to the central frequency (see Fig.~\ref{pulse}), they can give the dominant contribution to the pair production probability because the latter increases with increasing positron momentum [see Eqs.~\eqref{R_Taylor_1a}-\eqref{R_Taylor_1c}]. 

Thus, the availability of photons with relatively high energies in the short laser pulse, in combination with the {\it energy} dependence of the production probability, leads to the enhanced spin sensitivity observed in the present scenario.

\begin{figure}
 \includegraphics[width=0.85\columnwidth]{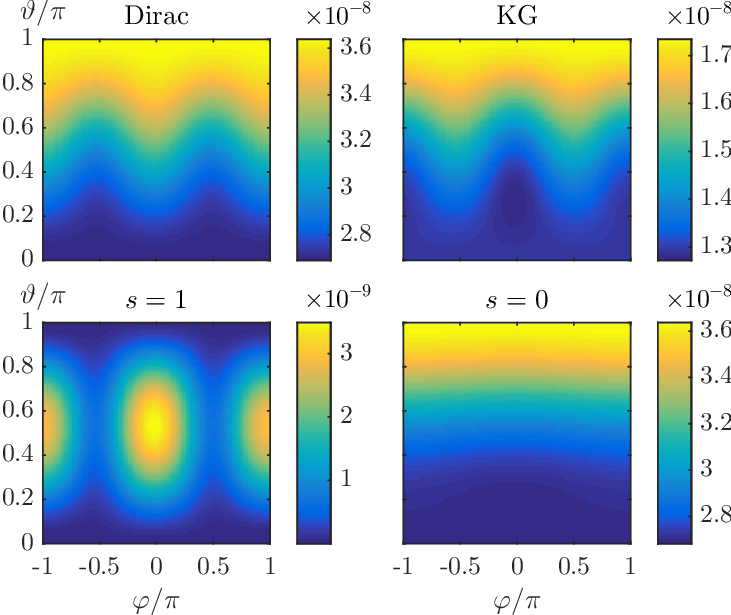}
\caption{Fully differential probability $\frac{d^3\mathscr{P}}{dE_{p_+}d^2\Omega_{p_+}}$ in units of $1/m$ for $p_+=0.15m$, $\omega_c=\omega_\gamma=1.006m$, $\Nosc=2$, and $\xim=0.02$. The angles $\vartheta$ and $\varphi$ describe the positron emission direction and are measured with respect to the laser propagation and polarization directions, respectively. }
\label{above-angles}
\end{figure}

While the difference between KG and Dirac particles in terms of their total production probabilities is amplified in the short laser pulse, certain similarities may still exist. This can already be anticipated by comparing the curves for KG and ``spinless'' Dirac pairs in Fig.~\ref{above-energy}, whose shapes resemble each other. Further insights can be gained from an inspection of the angular distributions of positrons generated in the short pulse. In Fig.~\ref{above-angles}, we present our corresponding results for positrons with fixed momentum $p_+=0.15m$ for different spin configurations. The production probability of Dirac pairs is mainly determined by the contributions with $s=0$ (see also Fig.~\ref{above-energy}). The corresponding angular distribution is almost homogeneous in the transverse direction and moderately enhanced in the laser backward direction \footnote{This asymmetry in the differential probability between the laser forward and backward direction is essentially caused by the asymmetry between the gamma quantum and the laser pulse. While the pulse delivers photons with a continuous momentum distribution, the gamma quantum has a fixed momentum.}. 
In contrast, the contribution from $s=1$ is suppressed and its angular distribution looks qualitatively different. Emission along the laser polarization axis is preferred, whereas emission along the collision axis is further suppressed. 
The comparison between the two Dirac contributions nicely illustrates the consequences of the conservation of angular momentum (see Sec.~III).
The full Dirac distribution is mainly determined by the contribution from $s=0$. The much weaker probability for $s=1$ appears as a modulation along the azimuthal direction. 
In comparison, the KG distribution shows certain similarities with the spin-summed Dirac distribution because it quite closely resembles the dominant case $s=0$. Also azimuthal modulations exist which, however, are offset by $\pi/2$ as compared with the full Dirac distribution. In Sec.~IV.C we shall see that the similarity between the KG and spinless Dirac distributions can be even more obvious.

\subsection{Just below the one-photon threshold}
As a second example, we choose $\omega_c=0.7 m$ and $\omega_\gamma=1.4 m$. Hence, the absorption of two laser photons is required in the monochromatic case. For $\xi=0.2$, we find strong spin effects with $\DiKg\approx5.8$. Conversely, for a short pulse with $\Nosc=2$, we find $\DiKg\approx3.4$.

\begin{figure}
 \includegraphics[width=\columnwidth]{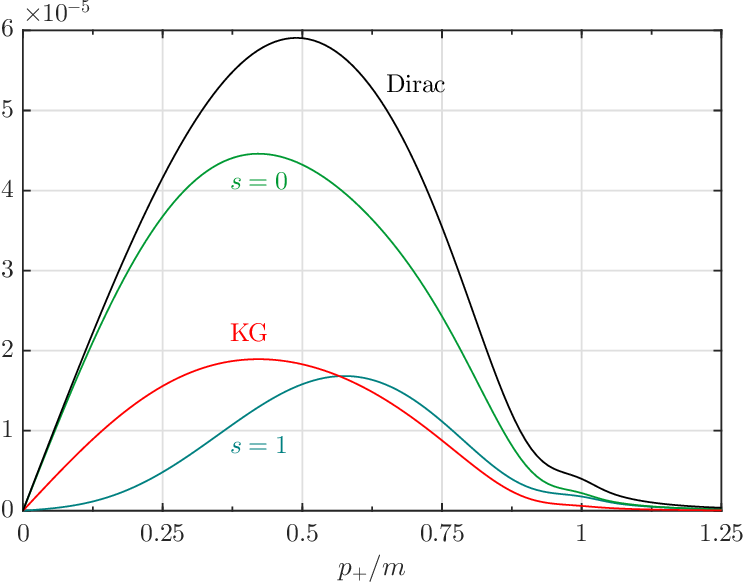}
\caption{Same as Fig.~3 but for $\omega_\gamma=1.4m$, $\omega_c=0.7m$, and $\xim=0.2$.}
\label{just-energy}
\end{figure}

In this case, the spin effects in the pulse are reduced as compared to the monochromatic field. Since the photon frequencies are just slightly below the one-photon threshold, the two-photon process in the monochromatic field leads to a pair with rather large value of $\beta\approx0.70$ (when dressing effects are neglected). From Fig.~\ref{zeta} we can confirm $\DiKg\approx5.9$, accordingly. When, instead, the pairs are created in the short laser pulse, the broad pulse spectrum contains photons with sufficiently large energy in order to drive one-photon processes. As depicted in Fig.~\ref{just-energy}, these one-photon processes produce pairs with lower values of $\beta$, such that the resulting spin effects are smaller than in the monochromatic field. In fact, the dominant contribution to the total creation probability stems from processes with positron momenta $p_+\approx 0.59m$, corresponding to $\beta\approx 0.51$. 

The dominance of the one-photon channels in the pulsed field is due to the $\xi^{2N}$ scaling of the process probability in the regime where $\xi\ll 1$. Therefore, pair production by two-photon absorption is suppressed by an additional factor of $\xi^2$. This phenomenon has been called ``subthreshold enhancement'' of pair creation in Ref. \cite{Kaempfer}, in analogy with the terminology in nuclear collisions. 

The reduction of the spin sensitivity in the present scenario thus results from the availability of photons with relatively high energies in the short laser pulse, in combination with the {\it intensity} dependence of the production probability.

We further note that, as before, the curves for KG and spinless Dirac pairs in Fig.~\ref{just-energy} have similar shapes. 
Besides, we point out that, for $p_+\gtrsim m$, the two-photon process becomes dominant also in the pulsed field and leads to the small plateau in the energy spectra of Fig.~\ref{just-energy}.

\subsection{Deeply below the one-photon threshold}
Finally, we consider a scenario where the reduction of spin effects in a short laser pulse is particularly pronounced. To this end, we investigate the case of $\xi=0.001$, $\omega_\gamma=1.006m$, and $\omega_c=0.503m$, which is half of the frequency of the above-threshold case in Sec.~IV.A. For the monochromatic field, these parameters lie deeply below the one-photon threshold. As a consequence, the two-photon process leads to a low-energy pair with $\beta \approx 0.11$ and very strong spin effects with $\DiKg \approx 39$. In comparison, the ultrashort pulse with $\Nosc=2$ leads to $\DiKg\approx2.9$ only, which means that the relevance of the spin degree of freedom is suppressed by an order of magnitude.

Based on the line of argument in Sec.~IV.B, one might suspect that the pronounced suppression comes from the fact that, in the short pulse, the pairs can be generated by a single energetic photon whose frequency equals $2\omega_c=1.006m$. However, a closer look at the spectrum of the pulse is necessary. It exhibits characteristic holes at (most) integer multiples of the basic frequency (see Fig.~\ref{pulse}). They are induced by the window function $\cf(\pv)$ in Eq.~\eqref{LasVecPot}. In particular, the pulse under investigation does not contain photons of frequency $2\omega_c$. As a consequence, the corresponding energy can only be provided by two (or more) photons. This means that the pulse produces pairs with $\beta\approx0.11$ predominantly through a two-photon process (just as the monochromatic field). This production channel is associated with a large value of $\zeta$.

\begin{figure}
 \includegraphics[width=\columnwidth]{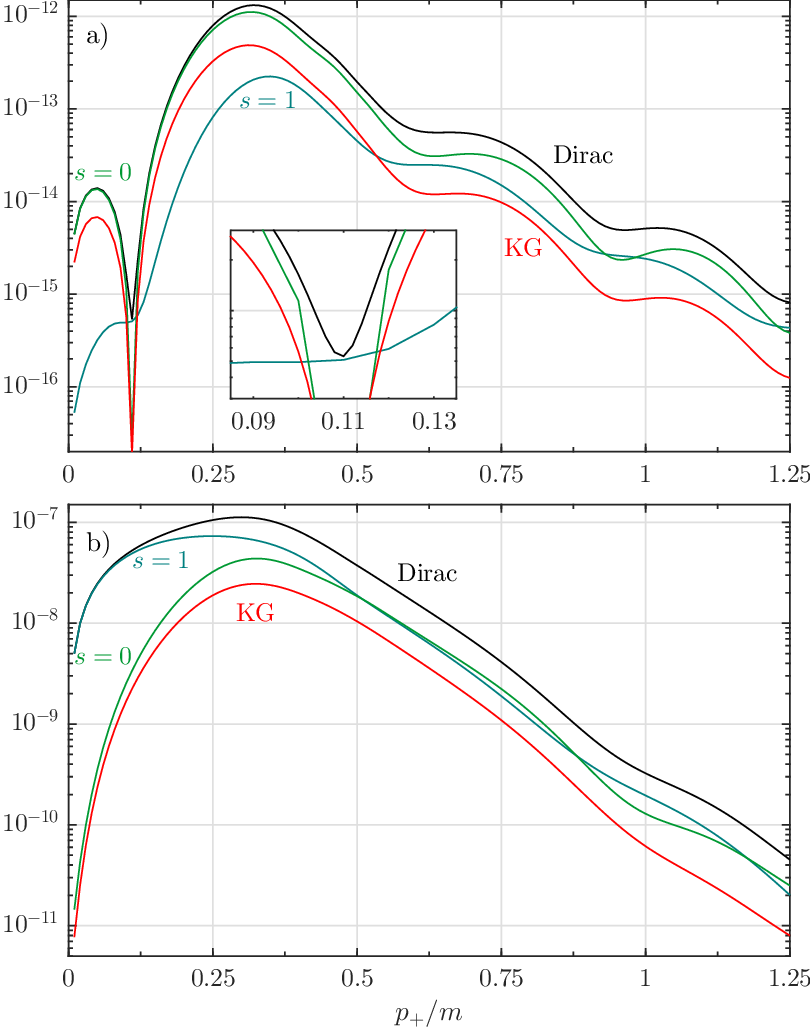}
\caption{Same as Fig.~3 but for $\omega_\gamma=1.006m$, $\omega_c=0.503m$, and a) $\xim=0.001$, b) $\xim=0.1$.}
\label{deep-energy}
\end{figure}

Additionally, as Fig.~\ref{deep-energy}a shows, the pulse produces pairs with other values of $\beta$ by means of one-photon processes. Due to the very small value of $\xim$, these one-photon processes deliver the dominant contribution to the full probability (see Sec.~IV.B). The typical positron momenta amount to roughly $p_+\approx 0.33m$, corresponding to $\beta\approx0.31$. Therefore, as the blue $\zeta_1$ curve in Fig.~\ref{zeta} indicates, the spin dependence of the total production probability in the pulse is heavily reduced as compared to the monochromatic field. It is interesting that in the scenario of Sec.~IV.A, the contribution of higher particle energies led to enhanced spin effects in the pulsed field, whereas in the current situation higher particle energies are accompanied by reduced spin sensitivity.

The reduction is particularly strong in the present case because the monochromatic field probes production of low-energy pairs by two-photon absorption which is highly spin sensitive (see green $\zeta_2$ curve in Fig.~\ref{zeta}). In contrast, pairs are mainly generated with intermediate energies through one-photon processes in the pulse. This phenomenon occurs despite the spectral holes of the specific laser pulse under consideration. It can therefore be expected to arise in pulses of other shapes, as well, whenever the corresponding process in a monochromatic field with $\xi\ll 1$ produces low-energy pairs by an even number of laser photons. 

Figure~\ref{deep-energy}a also shows that the contribution from $s=1$ to the production probability of Dirac pairs does not reveal the pronounced dip at $\beta\approx0.11$. As we have seen in Sec. III, the contribution from $s=1$ to the production of low-energy pairs is suppressed when only one laser photon participates. The pulse, however, allows to deliver the required energy through two photons. These processes do not suffer from the spectral hole which affects the one-photon processes.

The plateaus visible in Fig. ~\ref{deep-energy}a are related to the spectral holes. The production of a positron of momentum $p_+$ requires an energy $E_L(p_+)$ to be absorbed from the pulse. This energy generally depends on the emission direction.
For the current parameters, only the process with $p_+\approx0.11m$ happens in a c.m. system, such that $E_L$ becomes angle-independent. Therefore, the hole in the pulse spectrum located at the frequency $1.006m$ directly translates to the dip in the momentum distribution at $p_+\approx0.11m$. For other values of $p_+$, the angle-integrated probabilities comprise an interval of energies $E_L(p_+)$. Therefore, an average over some spectral range is taken, washing out the sensitivity to the spectral holes. This holds especially for higher positron momenta, where the lower and upper boundary of the interval are shifted to higher energies. Furthermore, the $E_L$ interval becomes broader and soon comprises several spectral holes. The local maxima between the holes (see Fig.~\ref{pulse}) provide the dominant contribution to the integrated probabilities. Since the spectrum falls off for higher frequencies, the integrated probability drops significantly whenever the lower interval boundary passes one more spectral maximum. In between, the integrated probability is almost constant, leading to the plateau structure.

Finally, we increase the intensity from $\xim=0.001$ to $\xim=0.1$, which strongly enhances the relative importance of processes with higher photon numbers due to the intensity scaling.
Unlike in the scenarios presented before, low-energy particles can be expected (from the P-Model, see \cite{Jansen}) to be predominantly produced by two-photon processes, while the three-photon process becomes important for higher energies.
Inspecting the positron energy spectrum as depicted in Fig.~\ref{deep-energy}b, the production of low-energy particles is mainly determined by the contribution with $s=1$, while the production of ``spinless'' particles is strongly suppressed. These results are in accordance with the results for two-photon processes in Sec.~III, but they differ from the energy spectra of the other scenarios.
Furthermore, in comparison with the low-intensity case [see Fig.~\ref{deep-energy}a], the effects of the spectral holes are washed out due the convolution inherent to multiphoton processes.
The production probability of Klein-Gordon particles close to the threshold contains significant contributions from one- and, possibly, three-photon processes.

Despite the strong spin effects inherent to the two-photon process at low energies, the full probability reveals only moderate spin sensitivity with $\DiKg\approx4.6$.
The main contribution to the full probability in the pulse stems from processes with intermediate momenta $p_+\approx 0.34m$. As we have already seen in Fig.~\ref{zeta}, the spin sensitivity at those energies is much weaker than for processes close to the threshold.
In contrast, the monochromatic field predominantly produces low-energy particles via a two-photon process, which reveals very strong spin effects with $\DiKg_2\approx 66$.
In comparison with the low-intensity case (see beginning of Sec.~IV.D), the rather small reduction of the particles' momenta due to the laser dressing leads to a significant increase in $\DiKg_2$.
When the weaker yet noticeable three-photon process is included, the combined spin-effects in the monochromatic field amount to $\DiKg\approx43$ in total, exceeding the spin-sensitivity in the pulse by one order of magnitude.
The strong difference between the spin sensitivities in the pulse and in the monochromatic field thus persists despite the substantial increase in the laser intensity.

\begin{figure}
 \includegraphics[width=\columnwidth]{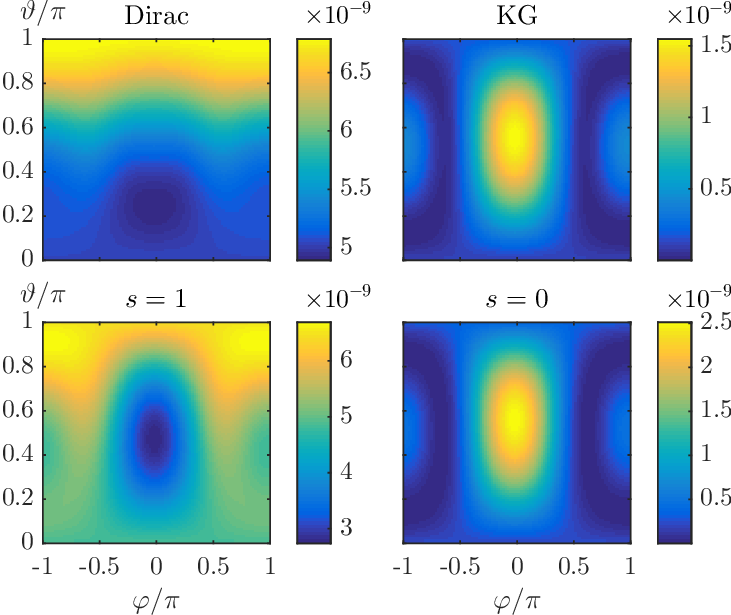}
\caption{Same as Fig.~4 but for $\omega_\gamma=1.006m$, $\omega_c=0.503m$, and $\xim=0.1$.}
\label{deep-angles}
\end{figure}

Figure \ref{deep-angles} shows a comparison of the angular distributions for fixed momentum $p_+=0.15m$. The results for Dirac and Klein-Gordon particles differ completely, but we find a striking similarity between KG and $s=0$. With the process being primarily induced by two laser photons, their numerical ratio is less than two (at least for the dominant directions).
In contrast to the angular distributions presented in Fig.~\ref{above-angles}, the similarity between the ``spinless'' channels occurs despite the fact that the contribution with $s=1$ is dominant.
We finally note that the strong localization of the ``spinless'' contributions is again a typical signature of the conservation of angular momentum in a two-photon process (see Sec.~III). The non-vanishing emission probabilities along the propagation direction are caused by processes involving different photon numbers.

\section{Conclusion}
We studied spin effects in strong-field Breit-Wheeler pair production in a short few-cycle laser pulse. This way, our previous investigations in \cite{KasiaBW,Jansen} were extended. Focussing on the regime of moderate laser intensities ($\xi\ll 1$ and $\xi\lesssim 1$), we compared predictions from laser-dressed scalar and spinor QED, based on the Klein-Gordon and Dirac equations, respectively. The results from spinor QED were also decomposed into spin contributions along the beam axis, which allowed us to obtain intuitive insights into the origin of the spin dependence of the pair production probabilities. 

We have found that the relevance of spin effects -- which was measured by the ratio $\zeta$ between total Dirac and Klein-Gordon probabilities -- can both be enhanced or reduced in a short laser pulse, as compared with a monochromatic laser field.  For instance, spin effects may be enlarged in the pulse when pairs are produced above the monochromatic one-photon threshold. In contrast, when low-energy pairs are produced by an even number of photons from a monochromatic field, the corresponding pair production process in the pulse will mainly occur with an odd number of laser photons, which can lead to a very strong suppression of the spin sensitivity. In particular, the value of $\zeta$ can either be much larger or significantly smaller than the value of $\zeta=6$, which is obtained in the strong-field tunneling regime of Breit-Wheeler pair production where $\xi\gg 1$ \cite{Selym}.

Besides, it was found that the predictions from the spinless Klein-Gordon theory often exhibit a close similarity with the ``spinless'' contribution from the Dirac calculation. This holds not only for total probabilities, but also for energy spectra and, most strikingly, angular distributions of produced positrons.

In conclusion, the comparability between scalar and spinor pair production strongly depends on the kinematical conditions.
The influence of a short pulse can be understood in terms of the properties of the production rates in a monochromatic field, combined with the frequency spectrum of the pulse.
The spin-sensitivity of the monochromatic rates reveals a strong dependence on the number of absorbed photons (especially its parity) and on the momentum of the particles.
The monochromatic field may probe channels with extremely small or large momenta, which may be either weakly ($\DiKg\approx2$) or strongly ($\DiKg\rightarrow\infty$) sensitive to the spin.
In contrast, when the comparison with a short pulse of the same central frequency is drawn, the full pair production probability in the pulse is mainly determined by production channels with intermediate particle energies -- which were found to reveal also an intermediate spin-sensitivity with $\DiKg\sim4$.

\section*{Acknowledgements}
Fruitful discussions with M. Dellweg are gratefully acknowledged.
This study was supported by SFB TR18 of the German Research Foundation (DFG) under Project No. B11 and the Polish National Science Center (NCN) under Grant No. 2014/15/B/ST2/02203. The exchange visits between the Universities of D\"usseldorf and Warsaw were sponsored by the the EU 7th Framework Programme under Grant No. 316244.

\end{document}